\documentclass[10pt,onecolumn,superscriptaddress]{revtex4-2}

\usepackage{graphicx,latexsym} 
\graphicspath{{images/}{../images/}}
\usepackage{amssymb,amsthm,amsmath,amsfonts}
\usepackage{bm}
\usepackage{dcolumn}
\usepackage{gensymb}
\usepackage{braket}
\usepackage{siunitx}
\usepackage{layouts}
\usepackage{lipsum}
\usepackage{float}
\usepackage{braket}

\newcommand{\AffCam}{Cavendish Laboratory, University of Cambridge, J. J. Thomson Avenue, Cambridge CB3 0HE, UK}
\newcommand{\AffOxEng}{Department of Engineering Science, University of Oxford, Parks Road, Oxford OX1 3PJ, UK}
\newcommand{\AffOxMat}{Department of Materials, University of Oxford, Parks Road, Oxford OX1 3PH, UK}
\newcommand{\AffCamEng}{Department of Engineering, University of Cambridge, Trumpington Street, Cambridge CB2 1PZ, UK}
\newcommand{\AffManPSI}{Photon Science Institute, Faculty of Science and Engineering, University of Manchester, Manchester M13 9PL, UK}
\newcommand{\AffManEE}{Department of Electrical and Electronic Engineering, Faculty of Science and Engineering, University of Manchester, Manchester M13 9PL, UK}

\begin{document}

\title{Laser Activation of Single  Group-IV Colour Centres in Diamond}

\author{Xingrui Cheng}
\affiliation{\AffOxMat}\affiliation{\AffOxEng}
\author{Andreas Thurn}
\affiliation{\AffCam}\affiliation{\AffOxEng}
\author{Guangzhao Chen}
\affiliation{\AffOxMat}
\author{Gareth S. Jones}
\affiliation{\AffOxMat}
\author{Maddison Coke}
\affiliation{\AffManPSI}
\author{Mason Adshead}
\affiliation{\AffManPSI}\affiliation{\AffManEE}
\author{Cathryn P. Michaels}
\affiliation{\AffCam}
\author{Osman Balci}
\affiliation{\AffCamEng}
\author{Andrea C. Ferrari}
\affiliation{\AffCamEng}
\author{Mete Atat\"ure}
\affiliation{\AffCam}
\author{Richard Curry}
\affiliation{\AffManPSI}\affiliation{\AffManEE}
\author{Jason M. Smith}
\email[Corresponding author: ]{jason.smith@materials.ox.ac.uk}
\affiliation{\AffOxMat}
\author{Patrick S. Salter}
\email[Corresponding author: ]{patrick.salter@eng.ox.ac.uk}
\affiliation{\AffOxEng}
\author{Dorian A. Gangloff}
\email[Corresponding author: ]{dag50@cam.ac.uk}
\affiliation{\AffCam}\affiliation{\AffOxEng}

\begin{abstract}
Spin-photon interfaces based on group-IV colour centres in diamond offer a promising platform for quantum networks. A key challenge in the field is realizing precise single-defect positioning and activation, which is crucial for scalable device fabrication. Here we address this problem by demonstrating a two-step fabrication method for tin vacancy (SnV$^-$) centres that uses site-controlled ion implantation followed by local femtosecond laser annealing with in-situ spectral monitoring. The ion implantation is performed with sub-50 nm resolution and a dosage that is controlled from hundreds of ions down to single ions per site, limited by Poissonian statistics. Using this approach, we successfully demonstrate site-selective creation and modification of single SnV$^-$ centres. The technique opens a window onto materials tuning at the single defect level, and provides new insight into defect structures and dynamics during the annealing process. While demonstrated for SnV$^-$ centres, this versatile approach can be readily generalized to other implanted colour centres in diamond and wide-bandgap materials.
\end{abstract}

\maketitle

\section{Main}\label{sec1}

Colour centres in diamond have garnered significant attention for their use in quantum technologies \cite{Hanson2008,Aharonovich2016} such as quantum simulators \cite{Randall2021}, quantum sensors \cite{Rondin_2014} and quantum networking interfaces \cite{Simon2010, Stas2022a}. Among these, Nitrogen-Vacancy (NV) centres are the most extensively studied due to their ground-state spin's long coherence times at room temperature \cite{Togan2010, Schroder:16}. However, their relatively low emission fraction into a purely photonic mode -- the zero-phonon line (ZPL) \cite{PhysRevX.7.031040} -- and susceptibility to spectral diffusion, particularly near surfaces \cite{Trusheim2020,Jelezko_2006}, poses a significant challenge for their use in optical quantum technologies. In contrast, group-IV colour centres have emerged as potential alternatives due to their crystallographic inversion symmetry, which leads to dominant ZPL emission \cite{Bradac2019}, manageable spectral diffusion near surfaces in nanophotonic devices \cite{ArjonaMartinez2022}, and reduced inhomogeneous broadening \cite{rogers_multiple_2014, Sipahigil_2016}, while preserving sufficient ground-state spin coherence at cryogenic temperatures \cite{Sukachev2017}. Tin-vacancy (SnV) centres stand out among group-IV defects due to their optimal spin-orbit coupling, which is significantly larger than for silicon-vacancy (SiV) \cite{Hepp_2014} and  germanium-vacancy (GeV) \cite{Bhaskar_2017} centres, protecting spin coherence against phonon scattering under standard cryogenic conditions ($\sim 2$K) \cite{Trusheim2020}, but smaller than for lead-vacancy (PbV) centres, allowing magnetically driven ground-state spin control with moderate strain levels \cite{Guo2023}. Despite these promising attributes, a method to create and activate single optically-active SnV (and other group-IV) centres that combines high spatial accuracy and high efficiency remains an outstanding challenge to scale up quantum technologies with these emitters.

Unlike nitrogen, group-IV elements are not naturally abundant in diamond. As a result, a number of different approaches have been developed to create group-IV colour centres, each with its own strengths and limitations. Chemical vapor deposition (CVD) growth~\cite{neu_single_2011, rogers_multiple_2014, iwasaki_germanium-vacancy_2015} and high-pressure, high-temperature (HPHT) synthesis~\cite{siyushev_optical_2017, palyanov_high-pressure_2019, ekimov_effect_2019} in the presence of group-IV precursors can produce high quality emitters. However, these methods lack the ability to position individual colour centres, which is crucial for many applications. In contrast, ion implantation enables high-resolution spatial positioning while allowing for fine control over the implantation dose~\cite{Adshead2023}, even facilitating the deterministic placement of individual ions~\cite{pacheco_ion_2017, smith_colour_2019}. However, particularly for large atoms like tin, this method often results in lattice damage that can degrade the optical and spin properties of the created emitters~\cite{iwasaki_germanium-vacancy_2015,Bradac2019, smith_colour_2019}. Post-implantation thermal annealing, crucial for creating the colour centres and repairing lattice damage, can be performed in two distinct ways. Low-pressure, low-temperature annealing at up to 1200~\textdegree C is compatible with most nanofabrication processes but often results in incomplete damage repair, resulting in significant inhomogeneous broadening of the ZPL distribution~\cite{Tchernij2017, rugar_characterization_2019, Gorlitz2020, Trusheim2020}. On the other hand, high-pressure, high-temperature (HPHT) annealing, at up to 2100~\textdegree C, can effectively reduce inhomogeneous broadening and produce colour centres with superior optical properties~\cite{Iwasaki2017,Gorlitz2020}. However, this process often causes significant damage to the diamond surface, making it incompatible with many nanofabrication techniques. These limitations have motivated the development of alternative approaches, such as shallow ion implantation and growth (SIIG)~\cite{rugar_generation_2020}. SIIG combines low-energy ion implantation with subsequent CVD overgrowth, minimizing lattice damage and thus allowing low-pressure, low-temperature annealing to sufficiently heal the diamond crystal structure \cite{rugar_generation_2020}. This method has shown promising results in creating high-quality, site-controlled SnV centres with low inhomogeneous broadening~\cite{rugar_generation_2020}. However, SIIG -- as well as the other approaches mentioned above -- still face challenges, including low formation yields (1\%-5\%)~\cite{Luhmann2019,rugar_generation_2020} and lack of site-specific control of the annealing process, which hinder scalable device fabrication and precise control over individual colour centre formation. These challenges are compounded by a lack of understanding of the defect structures formed by the implantation process, and the subsequent physical mechanism by which colour centre activation occurs during annealing.  

Laser writing has emerged as a favourable technique for precision engineering of colour centre defects in crystals through vacancy generation~\cite{Chen2017, Hadden2018, Day2023, wang2024} and laser annealing \cite{Chen_2019, Engel_2023}, and has made possible the deterministic creation of NV centres in diamond through live fluorescence monitoring and feedback\cite{Chen_2019}. Recent investigations have also shown that laser annealing can enhance the creation yield of silicon-vacancies during thermal annealing \cite{Tzeng2024}. These findings suggest the applicability of laser-based techniques to a broader set of emitters in diamond.

Here we report the laser-processing of ion-implanted electronic-grade diamond to create single SnV emitters at precise positions. Our fabrication path combines site-selective ion implantation, featuring 50 nm resolution (with sub-20 nm possible) \cite{Adshead2023}, with subsequent femtosecond laser annealing and live photoluminescence (PL) monitoring. By monitoring the emission spectrum during the anneal process, we observe switching between SnV$^-$ and another distinct Sn-related defect that had been observed previously but not directly associated with SnV \cite{Wahl2020,Corte_2022,Iwasaki2017}. This defect -- which we refer to as 'Type II Sn' and hypothesise to be a SnV defect bound to a carbon self-interstitial (SnV-C$_i$) -- is observed to be dominant post laser activation and appears to serve as a precursor state before a stable SnV$^-$ is accessed. Live spectral evolution data during the annealing process reveals reversible transitions between the Type II Sn centres, SnV$^-$ centres, and optically inactive states, providing a window into the defect formation process and making possible deterministic activation by PL-based feedback.

\subsection{Laser activation of defects in ion-implanted diamond}\label{subsec2.1}

\begin{figure}
    \centering
    \includegraphics[width=0.9\textwidth]{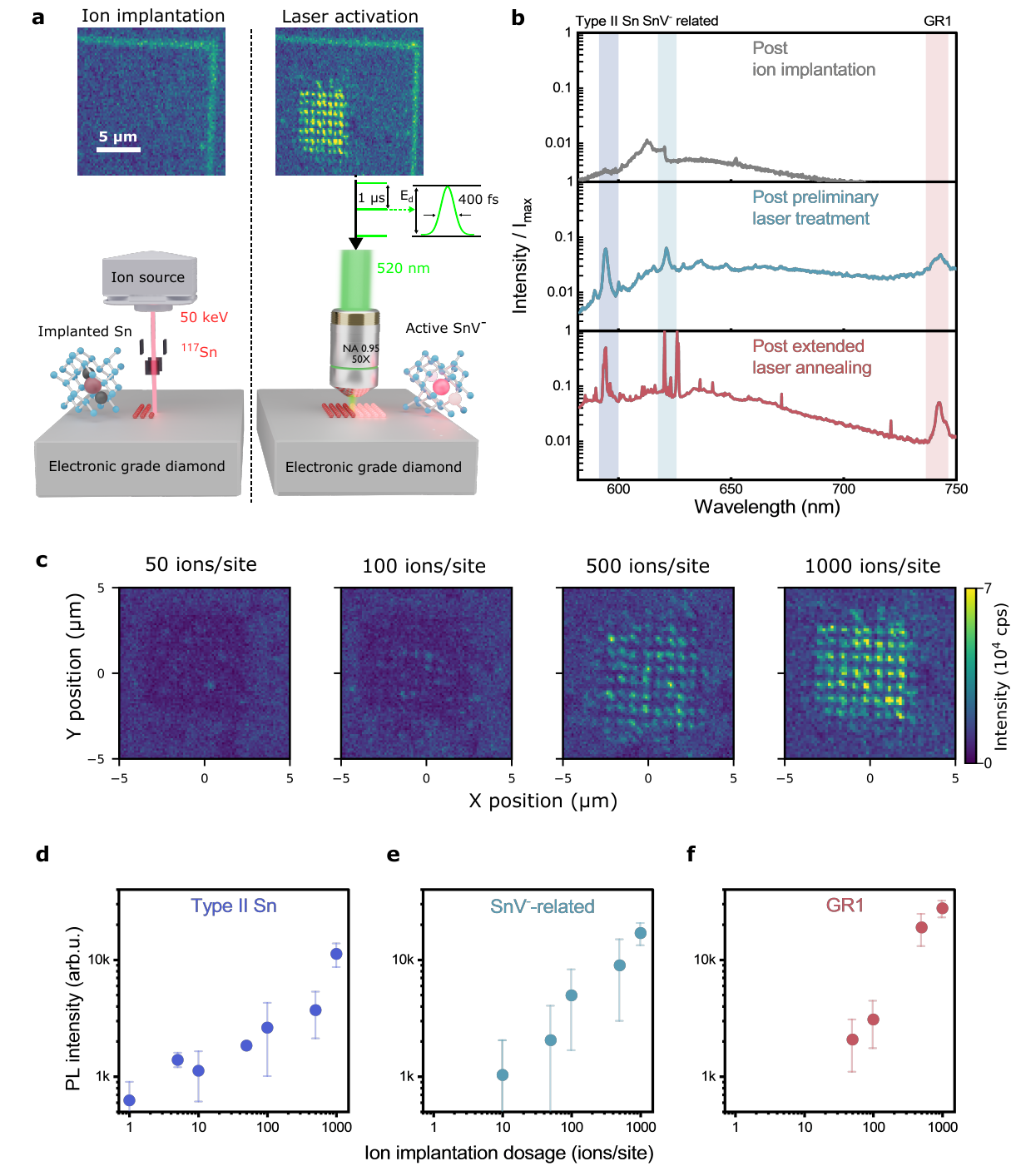}
    \caption{\textbf{Laser activation of Sn-related defect centres.} \textbf{a}, Schematic illustrating the process of ion implantation followed by femtosecond laser annealing. Sn$^{117}$ ions are implanted into the diamond lattice and, as a side effect, create lattice damage in the form of carbon vacancies and self-interstitials. Subsequent laser treatment activates SnV$^{-}$ and other defect centres. The two insets show 2D PL images with fluorescent alignment markers before and after preliminary laser treatment, respectively. The measurement was performed at room temperature using only a 532 nm notch filter to block the PL excitation laser.
    \textbf{b}, Example spectra from an implantation site with 500 ions at three different stages: immediately after ion implantation (top), after preliminary laser treatment (middle), and after 2 hours of laser 
    annealing (bottom). Highlighted areas are the spectral windows of interest for Type II Sn (blue), SnV$^-$ (green), and GR1 (red). $I_\text{max}$ is the peak spectral intensity from the bottom spectrum. 
    \textbf{c}, PL maps of four regions with different ion implantation dosages after preliminary laser treatment. The spectral collection window is $615-625$~nm.
    \textbf{d}, \textbf{e}, and \textbf{f}, Spectrally integrated emission intensity of Type II Sn-related, SnV$^-$-related, and GR1-related defects, respectively, as a function of implantation dosage following the preliminary laser treatment.}
    \label{fig:1}
\end{figure}

An illustration of the implantation and laser anneal process is shown in Fig.~\ref{fig:1}a. Doubly-ionised $^{117}$Sn atoms were implanted into a CVD grown Type IIa electronic grade diamond substrate (with a nitrogen density $< 1\,\text{ppb}$) using a focused ion beam \cite{Adshead2023} with an acceleration energy of 50 keV and a calculated average penetration depth of $\sim 20$ nm (see Methods). Multiple rectangular 100~\textmu m $\times$ 100~\textmu m arrays of implantation sites with 0.78~\textmu m spacing were fabricated. Arrays were implanted with a Poissonian mean number of 1000, 500, 100, 50, 10, 5, and 1 Sn ion(s) per implantation site. 

Following implantation, PL maps were obtained prior to femtosecond laser treatment by scanning a 1 mW, 532 nm continuous-wave excitation laser over the regions of interest (see Methods section). Fig.~\ref{fig:1}a (left panel) shows a typical post-implantation PL intensity map in which emission was collected in the wavelength range 550~nm to 800~nm revealing no visible fluorescence from the implantation sites. An absence of visible post-implantation fluorescence was observed even within the highest dosage (1000 ions/site) implanted region. A typical spectrum post-ion implantation is shown in Fig.~\ref{fig:1}b (grey curve), where the only notable feature is second-order Raman scattering of the excitation laser.

Laser annealing was performed with 400 fs pulses at a wavelength of 520 nm and a repetition rate of 1 MHz, focused at the surface of the diamond. As a preliminary activation step, the diamond was subjected to a raster scan of the activation laser across 5~\textmu m x 5~\textmu m regions within an implanted array (approximately 30 implantation sites) for each implantation dose, with a dwell time of about 1 second per site, at a pulse energy of 1.7~nJ (see Methods section). PL imaging following this preliminary step reveals fluorescent emission from the implantation sites, as shown in Fig.~\ref{fig:1}a (right panel). We note that the preliminary activation  laser treatment is well localised to the diffraction-limited laser spot (diameter 330~nm), as seen in Fig~\ref{fig:1}(c).  

Image analysis confirms the positioning accuracy of the sites to be sub-50 nm (see SI section 2). Spectral analysis of the observed emission is presented in Fig.~\ref{fig:1}b (blue curve) and reveals three predominant features absent in the spectra immediately after ion implantation. The first is an emission peak centred at 595~nm, which is known to be associated with Sn-related defects in low-temperature annealed diamond \cite{Iwasaki2017} and which we refer to here as Type II Sn. The second emission peak, centred at 620 nm, is characteristic of the SnV$^-$ defect, and a third peak at 740 nm is attributable to the well-known neutral vacancy (GR1)~\cite{DAVIES1977}. To isolate the activation of SnV$^-$ centres we recorded PL images recording only emission between wavelengths of 615~nm and 625~nm~\cite{Trusheim2020}. Fig.~\ref{fig:1}c shows PL intensity maps within this SnV$^-$ window for a range of implantation densities, revealing occasional, weak fluorescence from sites with 50 implanted ions extending to stronger fluorescence, from all sites, with 1000 implanted ions. Fig.~\ref{fig:1}d, e and f plot the relationships between implantation dosage and the PL intensities after initial laser activation, summed across the treated array sites, for spectral windows corresponding to Type II Sn, SnV$^-$, and GR1 defects (as labelled in Fig.~\ref{fig:1}b). As anticipated, increasing implantation dosage results in increasing PL emission intensities across all three spectral windows. Interestingly, however, only the GR1 fluorescence intensity grows linearly with implantation dose -- the increase in fluorescence from the Type II Sn and SnV$^-$ with ion dose are sub-linear, suggesting that larger doses lead to lower fluorescence quantum yields post-activation or a wider range of defect structures being formed.   

The SnV$^-$-region of the spectrum shown in Fig.~\ref{fig:1}b (blue curve) resembles the inhomogeneously broadened spectrum of an ensemble of SnV$^-$ centres formed by a thermal anneal around 800~\textdegree C (see SI of Iwasaki et al. ~\cite{Iwasaki2017}), suggesting only partial annealing of the diamond lattice with this preliminary laser treatment. To address this, we performed an extended laser treatment over a 2-hour period, focusing 1 nJ laser pulses to a single point in the diamond near to the activated array to ensure that no additional damage was created at the implantation sites of interest (see Methods section) -- note that this long exposure time reduces to only an effective 3 milliseconds of actual exposure with 7.2J of energy deposited.  This extended laser treatment was found to cause effects delocalised to several micrometres beyond the diffraction-limited laser spot (see SI section 1, including a supplementary study of the effect of exposure duration and laser pulse energy), and resulted in noticeable narrowing and strengthening of the fluorescence peaks, as shown in Fig.~\ref{fig:1}b (red curve). In particular, multiple sharp peaks are observed around 620 nm, suggestive of a population of distinct SnV$^-$ colour centres subject to different local strain environments. 

\subsection{Laser-activated single tin-vacancy centre}\label{subsec2.2}

\begin{figure}
\centering
\includegraphics[width=1\textwidth]{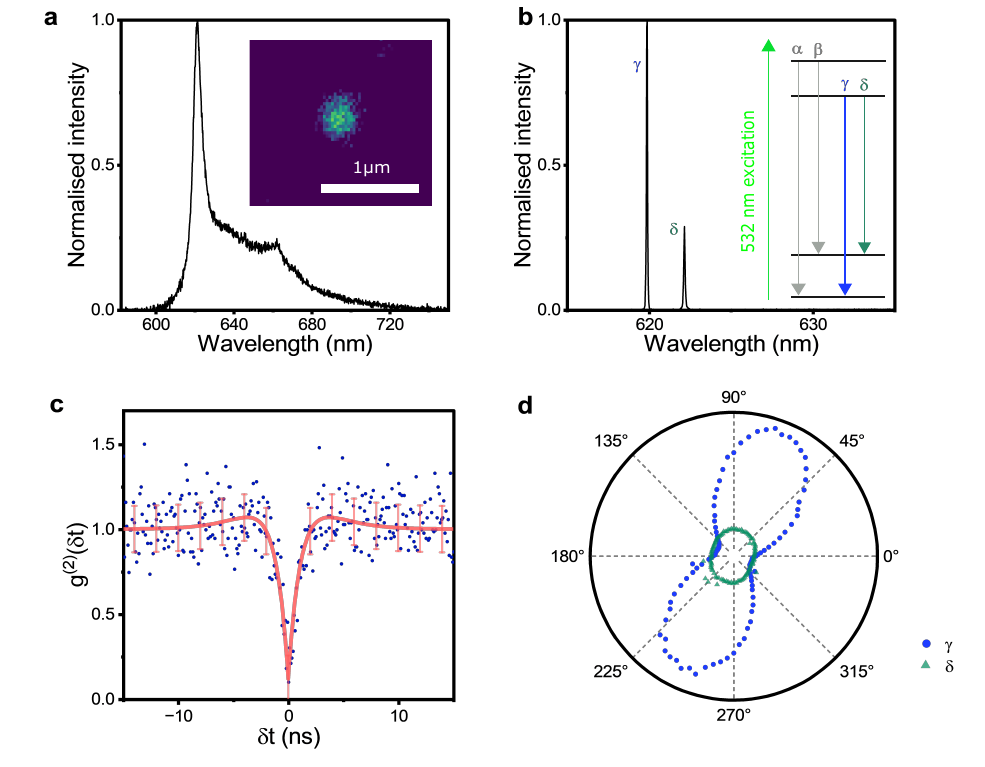}
\caption{\textbf{Characterization of a typical laser-activated single SnV$^-$ centre}. 
\textbf{a}, Room temperature PL spectrum of a site in the 10 implanted ions per site array, displaying a typical SnV$^-$ centre spectrum. The inset shows a PL intensity map of the site and its nearby surroundings.
\textbf{b}, Spectrum at 4.2~K of the same site, showing the $\gamma$ and $\delta$ optical transitions typical of an SnV$^-$ centre.
\textbf{c}, Histogram showing the background-corrected second-order auto-correlation measurement of PL emitted from the same site. Three-level model $g^{(2)}(\delta t) = 1 - (1 + \alpha) e^{-\frac{|\delta t|}{\tau_1}} + \alpha e^{-\frac{|\delta t|}{\tau_2}}$ (red curve), with fitted values $\tau_{1} = 1.4 \pm 1.5$~ns and $\tau_{2} = 9.7$~ns.
\textbf{d}, Polarisation dependence of the emission. Intensity of the $\gamma$ (blue dots) and $\delta$ (green triangles) transitions as a function of analysing polariser angle, normalized to the maximum value of each transition.}
\label{fig:2}
\end{figure}

By applying laser annealing to arrays containing an average of 10 ions per site, we now demonstrate the capability to laser-activate single implanted SnV$^-$ centres. Fig.~\ref{fig:2}a shows a room temperature PL spectrum (processed to remove Raman scattering) from one of the array sites, with the associated PL image in the inset. The spectrum reveals a remarkably clean ZPL at 619~nm and its accompanying phonon sideband (PSB) extending to 750~nm~\cite{Iwasaki2017}, characteristic of a single SnV$^-$ colour centre. At cryogenic temperature (4.2~K) the ZPL splits into two narrow peaks, characteristic of the \(\gamma\) and \(\delta\) optical transitions of the ZPL (see level diagram in Fig.~\ref{fig:2}b) \cite{Trusheim2020}. The splitting between the \(\gamma\) and \(\delta\) lines is 1.7~THz, suggesting that the SnV$^-$ centre is located in a highly strained environment\cite{Meesala2018}.

To confirm that the activated SnV$^-$ centre corresponds to a single emitter, we performed a Hanbury Brown-Twiss intensity auto-correlation measurement~\cite{HANBURYBROWN1956}. Fig.~\ref{fig:2}c shows the background-corrected \cite{Brouri:00} second-order autocorrelation function g$^{(2)}(\delta t)$, revealing a dip at $\delta t = 0$ with a g$^{(2)}$(0) value of \(0.1 \pm 0.13\), which is characteristic of a single quantum light emitter (see Methods section). Fitting of an analytic function for a three-level system \cite{Iwasaki2017} to the g$^{(2)}(\delta t)$ histogram (see SI) reveals an excited state lifetime of $ \sim $ \(1.4 \pm 1.5\)~ns . This value is shorter than the $5-7$~ns lifetimes reported previously for SnV$^-$   \cite{Trusheim2020,Gorlitz2020} which may be attributable to its close proximity to the diamond surface (20~nm depth) leading to rapid non-radiative decay \cite{Gorlitz2020}, delays of emission \cite{Eremchev2023}, or the effect of high power off-resonant excitation \cite{PhysRevLett.87.257401,Michler_2002}.

Fluorescence polarimetry (see Methods section), reveals a high degree of linear polarisation of the fluorescence from the $\gamma$ transition (Fig.~\ref{fig:2}d), also in line with previous reports~\cite{Tchernij2017, rugar_characterization_2019}.

\subsection{The Type II Sn defect complex}\label{subsec2.3}

\begin{figure}
\centering
\includegraphics[width=1\textwidth]{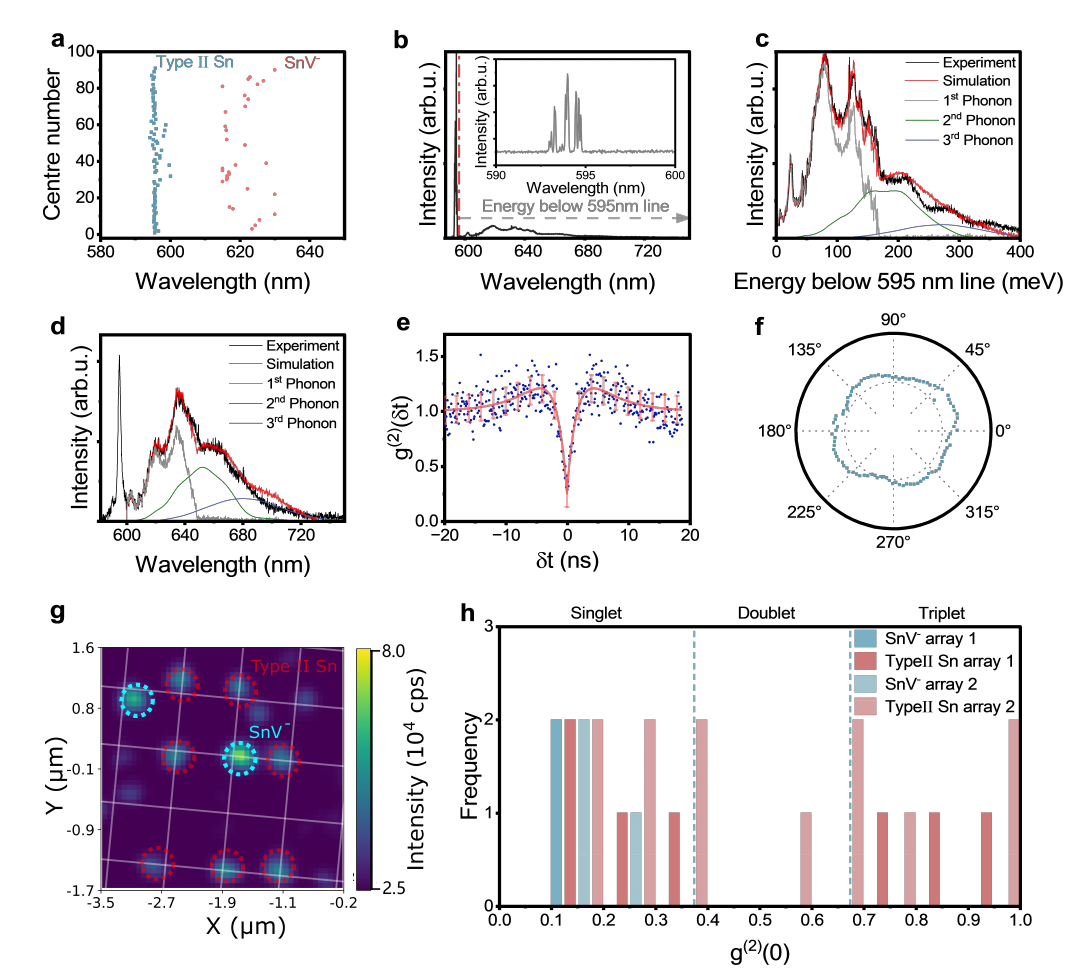}
\caption{\textbf{Study of Type II Sn defect complexes}.
\textbf{a}, ZPL distribution of 100 emitters activated by laser annealing.
\textbf{b} 4.2~K spectra of a single Type II Sn. The inset shows a high resolution spectrum of the ZPL (see also SI section 3).
\textbf{c} Analysis of the vibronic spectrum. The Type II Sn centre has sharp ZPLs at 595~nm and a PSB extending to 720~nm, which is decomposed into multi-phonon contributions.
\textbf{d} Room temperature spectrum of the same Type II Sn centre and analysis of the vibronic spectrum. 
\textbf{e}, Histogram displaying second-order autocorrelation measurement results with background correction and corresponding four-level fitting from the single Type II Sn centre. A clear dip appeared at 0 time delay with a g$^{(2)}$(0) value of 0.3.
\textbf{f}, Polarisation dependence of the Type II Sn ZPL emission measured at 4.2~K.
\textbf{g}, 2D PL map of an 3~\textmu m x 3~\textmu m array with a dosage of 10 ions/site, taken after 5 minutes of laser annealing. Nine stable emitters are identified, with SnV$^-$ centres circled in cyan and Type II Sn centres circled in red.
\textbf{h}, Histogram of g$^{(2)}$(0) values for activated stable emitters.}\label{fig:3}
\end{figure}

The appearance of a Sn-related defect complex with ZPL in the 595~nm spectral region -- which we refer to as 'Type II Sn' -- has been reported previously \cite{Wahl2020,Corte_2022,Iwasaki2017} but a detailed study has yet to be performed. Here we show the creation of single Type II Sn defects and analyse them in more detail, comparing them with the appearance of single SnV$^-$ colour centres.
Extended laser annealing (see Methods) of a 100-site array with an average of 10 implanted ions per site reveals the formation of sharp fluorescence peaks at both 595 nm and 620 nm, indicating the creation of small numbers of Type II Sn and SnV$^-$ colour centres. Fig.~\ref{fig:3}a shows the resulting distribution of wavelengths for the dominant (narrowest and most intense) spectral features at room temperature for these 100 activated sites. Interestingly, the Type II Sn centres at 595~nm exhibit a much narrower wavelength distribution, with standard deviation 1.10 nm, than the SnV$^-$ centred at 620 nm, with standard deviation 4.77 nm; this suggests differing strain susceptibilities and defect structures. The Type II Sn centres are also more numerous, with a ratio 7:3.

Fig.~\ref{fig:3}b, c, and d show high-resolution spectra of the Type II Sn fluorescence recorded at cryogenic and room temperatures, displaying a clear ZPL and red-shifted phonon sideband (PSB) extending up to 720~nm. Franck-Condon analysis (see Methods section) \cite{Smith_2011,GDavies1981}  of the phonon sideband provides a good fit based on single phonon coupling spectra extending to the 165 meV longitudinal optical phonon energy of diamond, and yield a Huang-Rhys factor of 1.70, significantly larger than the value of 0.57 for SnV$^-$~\cite{Trusheim2020}. 

Fig.~\ref{fig:3}e shows HBT measurements performed on a Type II Sn centre, yielding a g$^{(2)}$(0) value of \(0.3 \pm 0.12\), confirming single-photon emission, and a fitted lifetime of $ \sim $ \(2.2 \pm 0.3\)~ns. In contrast to the SnV$^-$ centre, the 595~nm ZPL emission from the Type II Sn defect shows no linear polarisation ( Fig.~\ref{fig:3}f). 

The observations above -- the lower inhomogeneous broadening, the different Huang-Rhys factor, and the unpolarised emission -- strongly suggest that the Type II Sn centre is an altogether different Sn complex to SnV$^-$, and not another optically active charge state of same defect. This is consistent with theoretical predictions which predict the ZPL of the optically-active neutral tin-vacancy centre to be at 681~nm \cite{Thiering2019}. 

Fig.~\ref{fig:3}g and h present yield statistics for the two types of emitters. Fig.~\ref{fig:3}g shows a PL map obtained after 5 minutes of laser annealing over a previously untreated section of the array, highlighting the sites where only Type II Sn emission, marked in red, and only SnV$^-$ emission, marked in cyan, is observed. Fig.~\ref{fig:3}h shows a histogram of g$^{(2)}(0)$ values measured across all sites in the activated array. This histogram indicates that of the sites in the array that show fluorescence, most are single photon emitters and that all five SnV$^-$-active sites are single photon emitters. While the total activation yield of implanted ions is still low (10 emitters out of an estimated 160 implanted ions), this relatively short laser anneal (which due to the low duty cycle of the pulse train in the annealing laser and rapid thermal diffusion in diamond equates to less than a second of energy transfer to the lattice) suggests that higher yields are possible, and the result clearly demonstrates the potential for site selective feedback. 

\subsection{Dynamic switching between single SnV$^-$ and Type II Sn colour centres}\label{subsec2.4}

\begin{figure}[H]
\centering
\includegraphics[width=0.8\textwidth]{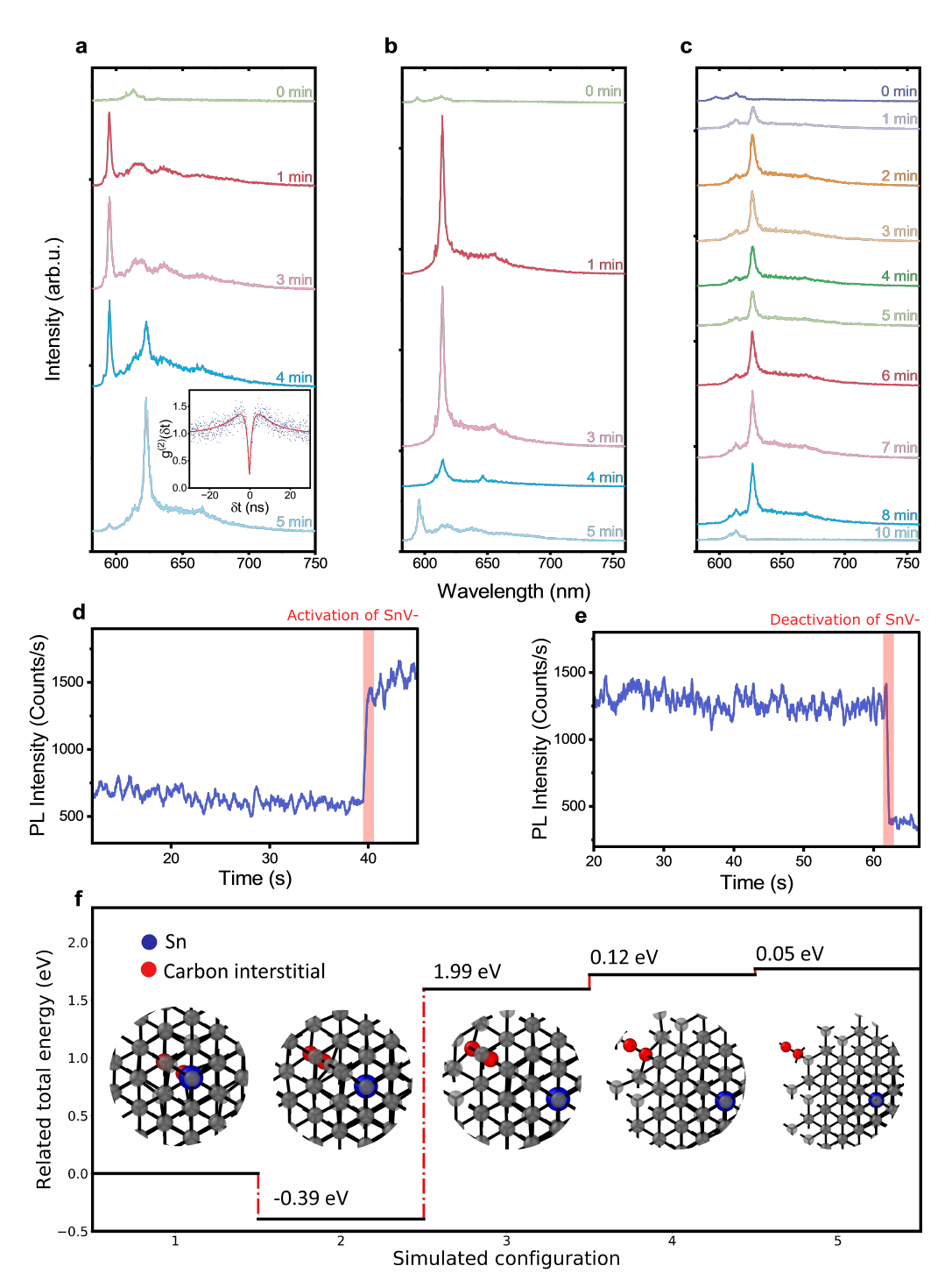}
\caption{\textbf{Live spectral monitoring of the temporal evolution of single SnV$^-$ centre during laser annealing}. Within each subplot, spectra progress from top to bottom, indicating increasing laser annealing time.
\textbf{a}, Formation of a single SnV$^-$ centre from a bare diamond substrate. The spectral evolution shows: the initial formation of an Type II Sn centre, the emergence of SnV$^-$ centre emission peaks, and the final SnV$^-$ centre spectrum. The inset shows a histogram of g$^{(2)}$($\delta$t) of the formed SnV$^-$ centre.
\textbf{b}, Formation of a single Type II Sn centre, the spectrum transitions from a Type II Sn centre (distorted emission) to a clear SnV$^-$ centre, terminating to a Type II Sn centre upon further laser annealing.
\textbf{c}, Deactivation of a SnV$^-$ centre. Starting with a Type II Sn centre (distorted emission), progresses to a clear SnV$^-$ centre. Further laser annealing first leads to modification of the SnV$^-$ emission, then deactivation into a bare diamond background.
\textbf{d} and \textbf{e}, SPAD monitoring during laser annealing with a collection window of $615-625$~nm to emphasize the signal from SnV$^-$ centres; showing a case of activation and deactivation of SnV$^-$ centres. 
\textbf{f}, Density functional theory (DFT) simulations of SnV$^-$-Ci complex with varying separation between the SnV$^-$ and Ci. The horizontal axis represents number of a lattice sites separation between the two defects. The two carbon atoms comprising the self-interstitial along [100] are indicated in red.
}\label{fig:4}
\end{figure}

Combining the laser annealing technique with PL spectroscopy facilitates the monitoring of dynamic processes at the single defect level. We select a previously untreated region of implanted array sites (with a mean dosage of 10 ions) to showcase this capability. Given the integration time of approximately a minute required for spectral measurements of single defects at ambient conditions, we adopted an iterative protocol alternating between laser annealing for 1 minute and the capture of PL maps for the identification of discrete emitters. Fig.~\ref{fig:4}a,b, and c each present a time-ordered sequence of PL spectra (without any background subtraction) for three selected single implantation sites, respectively, showing the evolution of the emission over several anneal steps.

Fig.~\ref{fig:4}a shows a sequence for a site where the spectrum is initially dominated by the second-order Raman signature of diamond, followed from 1 minute to 3 min of annealing by the appearance of Type II Sn fluorescence, and then between 4 min and 5 min a pregressive reduction in Type II Sn emission and appearance of SnV$^-$ emission. HBT measurement of this SnV$^-$ emission, displayed as an inset, reveals a g$^{(2)}$(0) value of 0.26 indicating a single colour centre. In the Fig. ~\ref{fig:4}b site the opposite process is observed: 1 minute of annealing results in strong  SnV$^-$ fluorescence which switches to Type II Sn fluorescence after 4 minutes of laser treatment. Fig.~\ref{fig:4}c describes yet another sequence of events for a site where SnV$^-$ emission  appears after one minute of annealing and survives for over 8 minutes of annealing, albeit with some modest variation in centre wavelength and intensity (see SI section 4). After 10 minutes, however, the SnV$^-$ emission disappears altogether, leaving only the Raman scattering signal. 

Simultaneously with the spectral monitoring, the switching dynamics of SnV$^-$ were probed with higher time resolution by monitoring the total fluorescence intensity within the $615-625$~nm emission window using a single-photon avalanche diode (SPAD). This method also provides faster, real-time feedback control, allowing annealing to be halted as soon as SnV$^-$ centres are activated or deactivated, similar to that used previously in the deterministic creation of NV centres ~\cite{Chen_2019}. Fig.~\ref{fig:4}d shows a SPAD trace during the 4-5 minute interval of Fig.~\ref{fig:4}a. After 40 seconds of laser annealing, a sudden increase in SPAD counts indicates the activation of the SnV$^-$ centre. Fig.~\ref{fig:4}e similarly illustrates the sudden deactivation of the SnV$^-$ centre during the 3-4 minute interval of Fig.~\ref{fig:4}b (see Supplementary Information).

\subsection{Discussion}

The fast, localised activation of fluorescence reported in Section \ref{sec1} suggests that it results from the direct optical excitation of pre-existing defects at the implantation sites. 
The high-energy Sn ion implantation process displaces some carbon atoms from their lattice sites, thereby generating both carbon vacancies and self-interstitials (Ci)~\cite{luhmann_screening_2018}. The absence of GR1 fluorescence in the post-implantation spectra suggests that these vacancies are predominantly in the negatively charged (ND1) state, which is not optically active under green excitation and does not emit in the measured wavelength range~\cite{kazuchits_luminescence_2022}. 
While unexpected for electronic grade diamond~\cite{davies_vacancy_1993, mainwood_stability_1997, isoya_epr_1992}, the formation of stable ND1 defects by the ion implantation process may be attributed to band bending effects near the surface~\cite{Wan_2020}, resulting from the relatively shallow implantation depth of approximately 20~nm. The preliminary laser treatment then converts the charge state of some of these ND1s to GR1s \cite{DAVIES1977, pu_negative_2001, kupriyanov_photochromic_2000, dannefaer_defects_2007, maki_time_2011}, resulting in the emergence of the corresponding luminescence feature centred at 740 nm.

The appearance of Type II Sn and SnV$^-$ fluorescence on short time scales may also be related to charge state modification, although this is not the hypothesis we deem most plausible based on our observations. Indeed, the femtosecond laser pulses trigger both linear and nonlinear absorption mechanisms at these implantation sites~\cite{Chen2017, Engel_2023}, creating free charge carriers and excitons via multi-photon band-to-band transitions and defect ionization~\cite{gattass_femtosecond_2008,Chen2017,PhysRevB.104.174303}. The resulting charge carriers can then transfer their energy to the diamond lattice through electron-phonon scattering and recombination processes~\cite{ichii_study_2020, PhysRevB.104.174303}. Our observation that the effects of extended laser treatment are delocalised to several micrometres beyond the diffraction-limited laser spot (SI section 1) suggest that such processes facilitate defect migration. As such some limited structural modification as a result of the preliminary laser treatment cannot be ruled out. In particular, carbon interstitials in diamond are highly mobile and may diffuse short distances as part of the activation process.

Regarding the formation of the 595~nm emission peak and SnV$^-$ centres, previous research has shown that $\sim 40\%$ of implanted Sn ions already settle into a split vacancy configuration \cite{Wahl2020}, necessarily resulting in two C$_i$ defects being present in the near vicinity. An alternate hypothesis is therefore that the resultant SnV-2C$_i$ complex is optically inactive, but that the initial laser treatment is sufficient to mobilise one of the C$_i$s, and could leave behind an optically active SnV-C$_i$ complex -- the Type II Sn with a feature at 595~nm. Importantly, subsequent extended laser treatment further facilitates the mobilization of the remaining $C_i$, which is more strongly bound to the SnV defect, allowing it to migrate away from the SnV defect and switching on the SnV$^-$ fluorescence. .

To investigate aspects of this hypothesis we performed density functional theory (DFT) calculations of the energy of a single C$_i$ near to an SnV$^-$ centre. The calculations reveal that a single C$_i$ can form a stable complex with SnV$^-$ without undergoing recombination. As can be seen in Fig.~\ref{fig:4}f, we find that the energy of the SnV$^-$ centre complexed with a C$_i$ one or two lattice sites away is around 2 eV lower than that of configurations where the carbon interstitial is either more distant from or directly adjacent (but not bound) to the Sn$^-$V centre.

The emerging picture of laser annealing mobilising C$_i$ defects to create different defect complexes is consistent with previous work on nitrogen-vacancy defects \cite{Kirkpatrick2024} and with other aspects of the experimental results presented here. The larger inhomogeneous distribution of spectral lines from SnV$^-$ than from Type II Sn, shown in Fig.~\ref{fig:3}a, would be a natural result of the varying strain field experienced by the SnV$^-$ defect as a result of the range of positions that the dissociated C$_i$ could occupy in the nearby lattice. Continued annealing leads to a clearer SnV$^-$ spectrum, as the C$_i$s diffuse away from the SnV$^-$ centres and creates a more stable environment that favours consistent emission characteristics. However, further annealing can also reintroduce the C$_i$s into the vicinity of the SnV$^-$ centres. In some cases, a C$_i$ will once again bind to the SnV$^-$ centre, producing Type II Sn emission or, if two Ci migrate close enough, deactivate the SnV$^-$ centres entirely (as seen in Fig.~\ref{fig:4}c).

If correct, our model may also shed light on the low yield of SnV$^-$ centres in undoped diamond, despite the high initial formation efficiency of split-vacancy configurations\cite{Tchernij2017,Wahl2020}. During the thermal annealing process, a significant proportion of SnV centres would likely remain bound to surrounding carbon interstitials, leading to the frequent formation of stable SnV-C$_i$ complexes emitting at 595~nm. Only if sufficient energy is provided to overcome the SnV-C$_i$ binding energy, by e.g. a 2000\textdegree C anneal (requiring high pressure), a large portion of SnV-C$_i$ will remain, as corroborated by previous work~\cite{Iwasaki2017}.


\section{Conclusion}\label{sec2}

In this study, we combined site-selective, low dose ion implantation and subsequent femtosecond laser annealing to activate  group-IV colour centres in electronic-grade diamond. Our approach allows for the activation of both ensemble and single SnV$^-$ centres, depending upon the ion implantation dose. This laser-based method offers a rapid and spatially selective approach to annealing under ambient conditions, eliminating the need for extreme high-pressure and high-temperature environments. Additionally, the technique allows for real-time spectral analysis and PL feedback control, providing unprecedented insight into the defect dynamics and making it an accessible method for on-demand activation with improved yields. By integrating low dose ion implantation with in-situ monitoring, deterministic activation of single group-IV defect centres is achieved. (Note that similar laser annealing of implanted Si ions can also be used to create isolated SiV$^-$ centres (see SI section 5)). Looking ahead, the technique offers potential for activation of deterministically implanted single ions, and for annealing with inline monitoring at cryogenic temperatures to enable in-situ tuning of a single defect's fine structure.

\section{Declarations}
\subsection{Acknowledgements}
This work was supported in part by the Quantum Computing and Simulation Hub (EP/T001062/1) through the Partnership Resource Fund PRF-09-I-06 (D.G., A.T.), by the Henry Royce Institute for Advanced Materials funded through EPSRC grants EP/R00661X/1, EP/S019367/1, EP/P025021/1 and EP/P025498/1, by the EPSRC through grants EP/R025576/1 and EP/V001914/1, and by capital investment by the University of Manchester (M.C., M.Ad., R.C.), by the EPSRC through grant EP/W025256/1 (P.S.), and by the ERC Advanced Grant PEDESTAL (884745) (C.P.M., M.At.). G.J. and C.P.M. acknowledge support from the EPSRC DTP. D.G. acknowledges support from a Royal Society University Research Fellowship.  We thank Element Six Ltd and Matthew Markham for providing electronic grade diamond for this project and IonOptika Ltd for discussions related to the Sn source.  


\subsection{Author Contributions}
X.C., A.T., G.J. and C.P.M. conducted the experiments and data analysis. G.C. performed the DFT calculations. M.C. and M.Ad. performed the ion implantation. O.B. processed the diamond sample. J.S., P.S., R.C., M.At. and D.G. conceived the experiments. J.S., P.S. and D.G. supervised the experiments. All authors contributed to the discussions of the data and to writing the manuscript.

\subsection{Data availability}

The data that support the findings of this study are available from the corresponding authors upon request.

\subsection{Competing interests}
The authors declare no competing interests.

\section{Methods}\label{sec4}

\subsection{Ion Implantation}

The diamond substrates were implanted using the Platform for Nanoscale Advanced Materials Engineering (P-NAME, University of Manchester, UK \cite{Adshead2023}), which has a maximum anode voltage of 25 kV. The isotopic mass resolution of the ion implantation system enables the selection of the $^117$Sn isotope, with choice of an ionisation state of 2+ providing an energy of 50 keV, leading to an implantation depth of approximately 20 nm (as simulated using Stopping Range of Ions in Matter, SRIM). The alignment markers, designed to be the vertices of a square, were created with 25 keV bismuth ions.
 
The implantations were conducted in a Poissonian mode. The ion beam current was kept fixed, and the pulse width of the electrostatic beam blanking was varied in order to obtain a desired average number of ions per spot (with the variance governed by Posissonian statistics).

The ion beam was electrostatically blanked between each implantation site, ensuring the only regions to be implanted were those chosen by the implantation design. The high vacuum of the sample chamber (10-8 mBar) ensured no contamination during the implantation process.

\subsection{Laser activation process}\label{subsec4.2}

The frequency-doubled output of an ytterbium-doped laser (Spectra Physics Spirit) was used to deliver a train of pulses with duration of 400~fs at a repetition rate of 1 MHz and a central wavelength of 520 nm. The laser beam was expanded and directed onto a Meadowlark liquid crystal spatial light modulator (SLM), which was in turn imaged in 4f configuration onto the back aperture of an Olympus 50$\times$ 0.95~NA objective lens. The SLM gave fine control over position and shape of the focal intensity distribution of the laser and was used to eliminate any system aberration.  

For the initial laser pre-treatment, the laser focus was uniformly scanned across ion implantation regions with varying dosages. A 5~\textmu m by 5~\textmu m area within the each array was raster scanned at a speed of 200 nm/s, and a laser pulse energy of 1.7~nJ. Given a diffraction limited laser focal size of 330~nm, this corresponds to an effective dose at each point in the raster scan of $1.6\times10^6$ pulses.

For the extended laser annealing, the laser pulses irradiated the diamond surface with a fixed focal position. The laser focus was placed approximately at the centre of an implantation region that we intend to study. The duration of the laser exposure was determined based on live PL feedback. The pulse energy used for this process was reduced to 1~nJ and the laser slightly defocused by 1~\textmu m from the diamond surface to avoid ablation during the extended exposure. 

 \subsection{Experimental confocal setup}\label{subsec4.2}

Room temperature and cryogenic PL imaging and spectroscopy were conducted using a home-built scanning confocal microscope integrated with the fs laser processing system described earlier. At room temperature, PL images and spectra were obtained using a 532~nm continuous wave (CW) laser (Cobalt), operating at a power of 1~mW as the excitation source. The PL was collected through a high numerical aperture (NA = 0.95) air objective lens, shared with the femtosecond laser activation beam. A tailored 582~nm long pass filter was applied when taking PL images to block the first order Raman emission of the diamond. Integrated PL measurements were made using a single photon avalance detector(SPAD) (Excelitas), while spectral data was acquired with a 300~mm spectragraph (SpectraPRo HRS-300) and low noise camera (PIXIS 100 from Princeton Instruments).

For cryogenic PL mapping, the sample was mounted in a Montana S-50 closed-cycle liquid helium cryostat, integrated with a customised beam-scanning confocal microscope. The sample was mounted upon a 4~mm$^{2}$ piece of thermal grade diamond using Apiezon thermal N-grease and silver dag. Additionally, the thermal grade diamond was thermally coupled to a cold finger with thermal N-grease. The cold-finger was connected to a three stage, open-loop piezoelectric actuator for navigation. Sample imaging was acquired with a Zeiss EC Epiplan 100X, 0.85 NA, 0.87 mm working distance objective lens. The objective lens was mounted within the cryostat housing under vacuum, and held at room temperature with an internal heater.

\subsection{Phonon autocorrelation measurements and data fitting}\label{subsec4.3}
The background-corrected \cite{Brouri:00} second-order autocorrelation function,  $g^{(2)}(\delta t)$ is defined by 

 \begin{equation}
g^{(2)}(\delta t) = \frac{\langle I(t)I(t + \delta t)\rangle}{\langle I(t) \rangle^2}
\label{eq1}
\end{equation}
where $I(t)$ is PL intensity at time difference of $(\delta t)$.

The fitting function for a three-level system \cite{Gorlitz2020} is defined by
\begin{equation}
g^{(2)}(\tau) = 1 - (1 + \alpha) e^{-\frac{|\tau|}{\tau_1}} + \alpha e^{-\frac{|\tau|}{\tau_2}}.\label{eq1}
\end{equation}
where \(\alpha\), \(\tau_1\) and \(\tau_2\) are all fitting parameters that relate to interlevel rate constants and the fitted \(\tau_1\) can be used to estimate the excited state lifetime of the emitter. The red sold lines in the autocorrelation plots in the main text are a least-squares fit using this equation.


\subsection{Polarisation Measurements}\label{subsec4.3}

To characterise the polarisation properties of the light emitted from the colour centres, we used a motorised half-wave plate (HWP) in the detection beam path. This HWP can be precisely rotated to vary the polarisation state of the transmitted light. The HWP introduces a phase shift between the orthogonal components of the light, effectively rotating its polarisation direction by twice the angle of the waveplate's rotation. Following the HWP, a linear polariser was positioned to transmit only the component of the electric field aligned with its transmission axis, thereby blocking the orthogonal component. The linearly polarised light was then coupled into a single-photon avalanche diode (SPAD) for intensity measurement. By rotating the HWP and recording the transmitted light intensity, we analysed the polarisation response of the SnV$^-$ centres using Malus’s Law, which relates the intensity of transmitted light to the angle between the light's polarisation direction and the polariser's axis.

The PL emission intensity was obtained by integrating the spectrum over the ZPL wavelength range. For Type II Sn defects, this range was $593-595$~nm. For the SnV$^-$ centre, the integration was performed over $619-622$~nm for the \(\gamma\) transition and $622-625$~nm for the \(\delta\) transition.

\subsection{Franck Condon Analysis}\label{subsec4.4}

To model the PL emission spectrum of Type II Sn, we applied the Franck–Condon principle \cite{Smith_2011, GDavies1981}. The emission intensity as a function of photon energy, \(I(E)\), is expressed as:

\begin{equation}
I(E) \propto E^3 \sum_{n=1}^{\infty} I_n(E_v) |M_{0n}|^2,
\end{equation}

where \(E\) represents the photon energy, \(I_n(E_v)\) is the lineshape corresponding to the creation of \(n\) phonons with vibrational energy \(E_v\), and \(|M_{0n}|^2\) is the squared overlap integral. This expression captures the sum of all possible phonon sidebands, each weighted by its Franck–Condon factor and modulated by the photon density of states.

The Franck–Condon factor quantifies the probability of emission into the \(n\)-phonon vibrational level of the ground electronic state, and is given by:

\begin{equation}
|M_{0n}|^2 \propto \frac{S^n}{n!},
\end{equation}

where \(S\) is the Huang–Rhys factor. This factor reflects the strength of the coupling between electronic states and vibrational modes, which in turn influences the relative intensity of the phonon sidebands.

The lineshape for the transition involving the creation of \(n\) phonons, \(I_n(E_v)\), is determined recursively through the following convolution of the single-phonon lineshape \(I_1(E_v)\):

\begin{equation}
I_n(E_v) = \int_0^{E_{\text{max}}} I_1(x) I_{n-1}(E_v - x) \, dx.
\end{equation}

Here, \(E_v\) denotes the vibrational energy of the lattice immediately following the optical transition, and \(E_{\text{max}}\) is the maximum energy of the n-phonon distribution. This recursive convolution process aggregates the contributions of successive phonon interactions, forming the overall lineshape for the phonon sideband.
The emission spectrum \(I(E)\) is a cumulative outcome of the weighted contributions from each phonon sideband. Each term in the sum reflects the probability of a specific phonon transition, as modulated by the Franck–Condon factors and the photon density of states.

\subsection{DFT model for SnV$^-$-Ci complex}
Simulations were performed using a 3 × 3 × 3 supercell containing a negatively charged tin-vacancy complex (SnV$^-$) and a carbon interstitial. When introducing one interstitial carbon into the diamond lattice, it undergoes energetic optimization to form a split-$\langle100\rangle$ interstitial carbon pair. Various configurations were explored with the interstitial placed at different lattice sites relative to the SnV$^-$ centre. These simulations indicate the presence of a stable configuration for the Type II Sn centre in which the interstitial does not recombine with the vacancies. The samples containing SnV and Ci are constructed through CrystalMaker \cite{palmer2015visualization} and then applied to DFT for geometry optimization based on the CASTEP code~\cite{clark2005first}. The Perdew-Burke-Ernzerhof (PBE) exchange-correlation functional \cite{perdew1996rationale} is utilised with ultrasoft pseudopotentials. The cutoff energy is set to 900 eV and the Monkhorst-Pack (MP) grid spacing is specified to 0.05.

\section{Supplementary information}

\subsection{Effect of Pulse energy and annealing time}
\begin{figure}[H]
\centering
\includegraphics[width=1\textwidth]{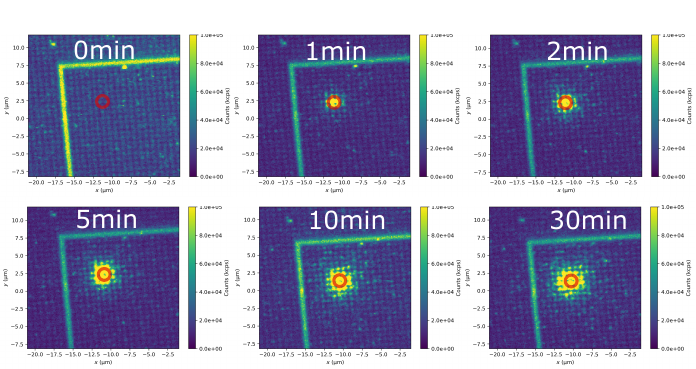}
\caption{\textbf{Study of the effects of laser annealing time}. Sets of 2D PL images are shown for a region implanted with 1000 ions per site, taken at the same locations after varying annealing times (annotated in white within each PL map). The laser was focused on the red circled points in the PL maps, with a slight defocus of 1 $\mu$m from the surface.}\label{P1 time}
\end{figure}

To gain deeper insight into the laser annealing process, we studied the effects of annealing laser pulse energy and annealing time on the activation of implanted emitters. Fig.~\ref{P1 time} presents a series of PL maps of the same site within a region implanted with 1000 ions per site. The site was annealed for 30 minutes with a pulse energy of 1.05 nJ, with the laser focused on the red circled points, which remained unchanged throughout the process. The annealing was divided into six distinct intervals, with PL images captured at each interval under identical experimental conditions. We note that in Fig.~\ref{P1 time}  the colour scale is set to the same upper limit in all plots to better visualise the radial spread of the activation, and thereby causing the PL intensity at the centre of the laser focus to appear saturated.

\begin{figure}[H]
\centering
\includegraphics[width=0.8\textwidth]{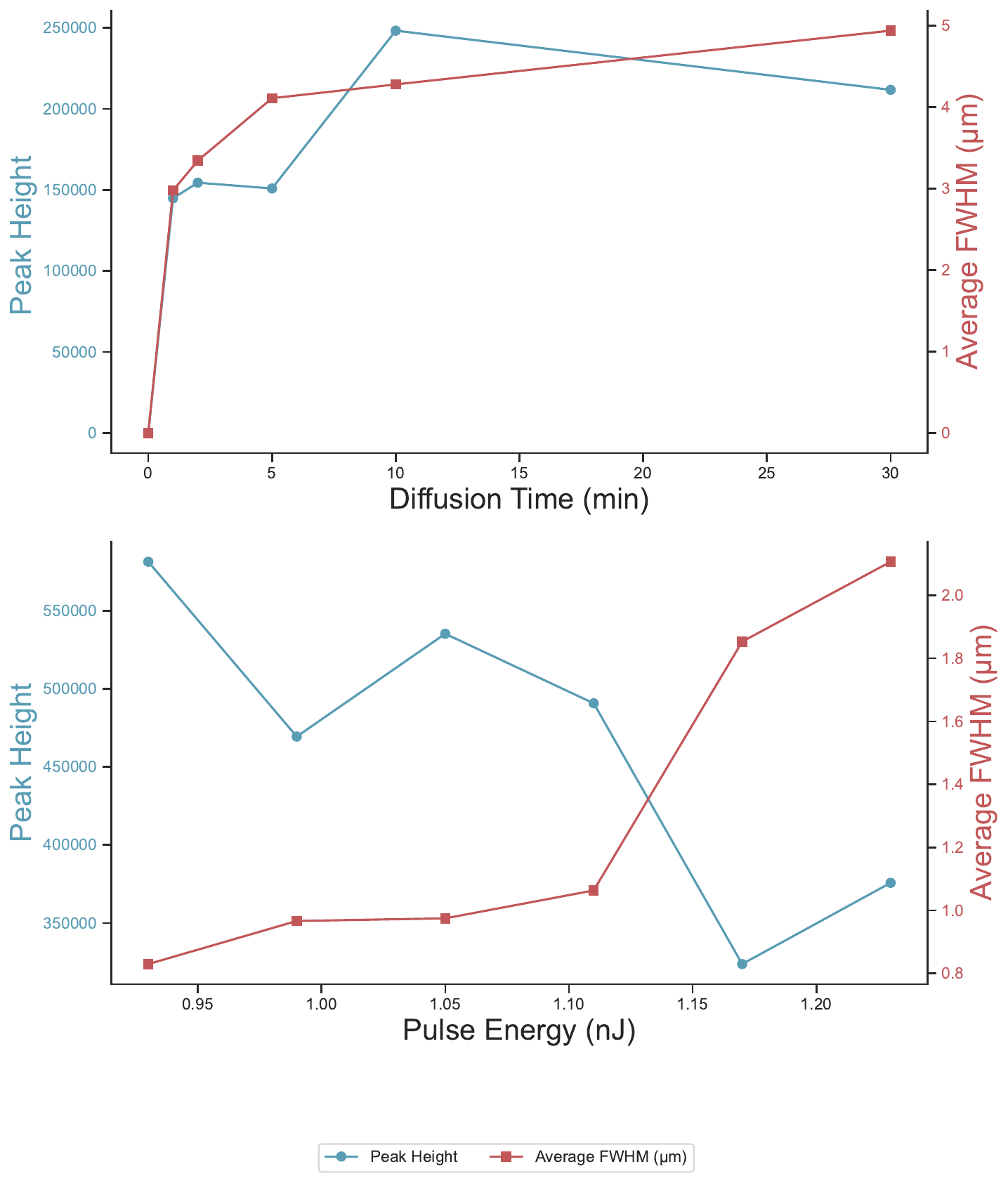}
\caption{\textbf{Average FWHM and Peak Height vs Pulse Energy and Diffusion Time}. The top graph shows the relationship between diffusion time and both the Average FWHM of the activated area and the peak height of the PL intensities for a region implanted with 1000 ions per site. The bottom graph displays the variation of the same parameters with respect to pulse energy. }\label{PulseE_Time}
\end{figure}

To quantify the spread and intensity of the activated area, a Gaussian function was fitted to the PL maps, treating PL intensity (here we are collecting emissions from both Sn-related defects and GR1s, and the figures displayed a min and max of brightness for better visibility, no capping was applied for any analyses) as the Gaussian height and analysing the spread across the \(x\) and \(y\) axes. The full width at half maximum (FWHM) and Gaussian peak height were extracted and plotted against annealing time, as shown in Fig.~\ref{PulseE_Time}.

The FWHM along the \(x\) and \(y\) axes was determined by:

\begin{equation}
\mathrm{FWHM}_{x/y} = 2\sqrt{2\ln2} \times \sigma_{x/y}.
\end{equation}

An average FWHM was calculated, and the peak height was taken as \(I\).

\begin{figure}[H]
\centering
\includegraphics[width=1\textwidth]{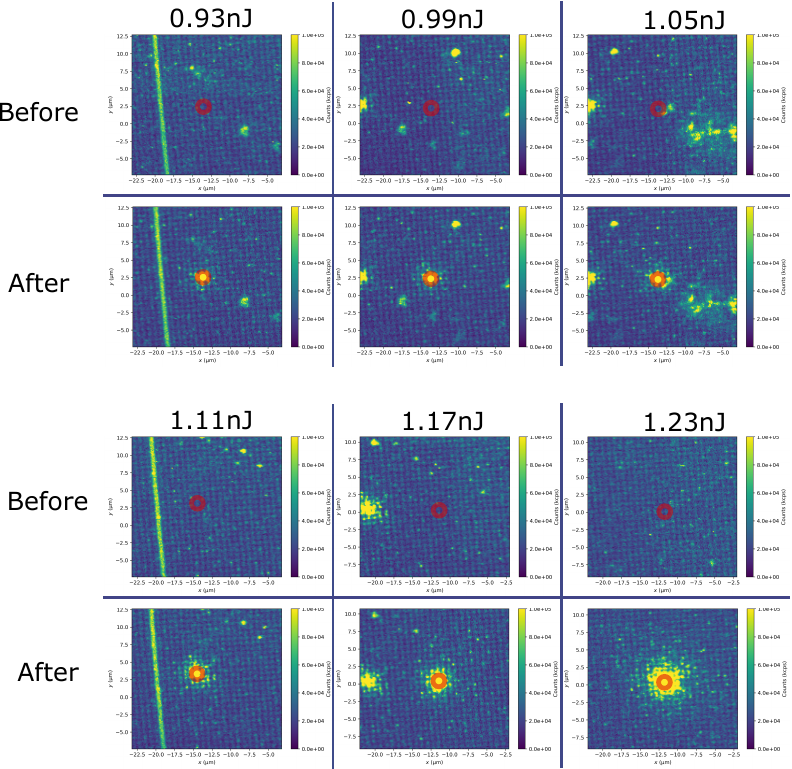}
\caption{\textbf{Study of the effects of annealing laser's pulse energy}. Sets of 2D PL images are presented for a region implanted with 1000 ions per site, captured before and after 1 min of laser annealing. The images progress from left to right and top to bottom, corresponding to increasing pulse energy. The laser was focused on the red circled points in the PL maps, with a  slight defocus of 1 $\mu$m from the surface.}\label{P1 pulseE}
\end{figure}

The effects of pulse energy were also investigated, with the annealing time held constant while varying the pulse energy of the annealing laser. This study was conducted in the region implanted with 1000 ions per site, where the laser was focused on the red circled points and annealed for 1 minute. Pulse energies ranging from 0.93 nJ to 1.23 nJ were applied. The PL maps were processed as described previously, and the variations in the spread and intensity of the activated region are presented in in Fig. \ref{PulseE_Time}.

Both the peak height and spread of the activated region increase with annealing time up to 10 minutes, after which they plateau. With a laser spot size of approximately 0.4~\textmu m, the activated region can extend up to 5~\textmu m, which is comparable to the exciton diffusion length in high-purity diamond at 298~K \cite{MORIMOTO201647}. This observation supports our hypothesis that the extended laser annealing process is exciton-mediated, where femtosecond laser pulses generate hot carriers that relax to form excitons \cite{PhysRevB.104.174303}. These excitons subsequently deliver energy to the lattice, facilitating the annealing process.

Increasing the pulse energy of the annealing laser has a minimal effect on the peak height, which is primarily constrained by the ion implantation dosage. However, higher pulse energy results in a slight increase in the spread of the activated region. Notably, at a pulse energy of 1.23 nJ, there is a significant drop in the fitted peak height and a marked increase in the fitted geometric mean. This behaviour is attributed to graphitisation of the diamond \cite{Chen2017}, where the diamond lattice locally melts and recrystallises, reducing the number of vacancies and interstitials, leading to the observed changes in the fitted data.

\subsection{Activated SnV$^-$ positioning accuracy}
A PL map of the diamond sample was analysed to quantify the positioning accuracy of the activated Sn-related centre array. Array with an implantation dose of 500 ions per site was selected for this study because the sites can be nearly fully activated by the femtosecond laser without the activated centres becoming too bright. The analysis focused on a central 5~\textmu m x 5~\textmu m area of the activated array. The colourbar limits were chosen such as to only show the implantation sites and not the background. Gaussian fits were applied to each individual emitter within the region of interest (ROI), with the centre of each fitted Gaussian being considered as the emitter's position.

\begin{figure}[H]
\centering
\includegraphics[width=1\textwidth]{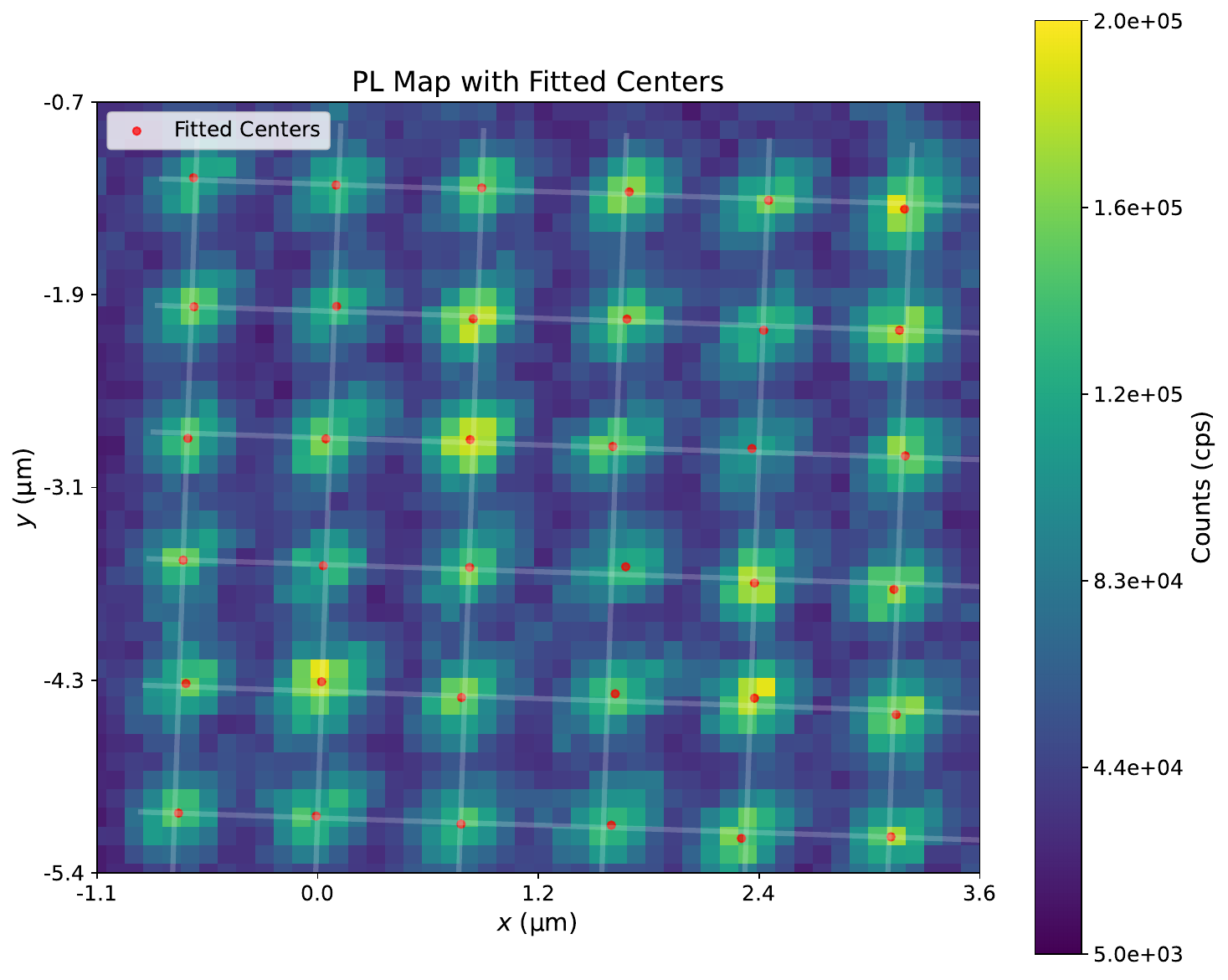}
\caption{\textbf{PL image of a post-laser activation Sn-implanted array}.
2D PL map of an array implanted with approximately 500 Sn ions per site, following laser activation. The white grid lines correspond to the theoretically fitted positions according to the implantation mask, and the red dots correspond to centre of the fitted Gaussian of individual emitter.}\label{P3 dis}
\end{figure}

\begin{equation}
f(x, y) = \text{offset} + A \times \exp\left(-\left(\frac{(x - x_0)^2}{2\sigma_x^2}\right) - \left(\frac{(y - y_0)^2}{2\sigma_y^2}\right)\right),
\end{equation}
where \textit{offset} is the constant background level, \(A\) is the peak height of the Gaussian, \(x_0\) and \(y_0\) are the coordinates of the centre of the Gaussian, and \(\sigma_x\) and \(\sigma_y\) are the standard deviations in the x and y directions, respectively.

An ideal grid with a known uniform spacing of 0.78~\textmu m, derived from the implantation mask, was fitted to the observed Sn implanted positions by optimising the x-offset \((dx)\), y-offset \((dy)\), and rotation angle \((\theta)\). The optimisation minimised the total discrepancy:

\begin{equation}
D = \sum_{i} w_i \min_j \sqrt{(x_{0,i} - x'_j)^2 + (y_{0,i} - y'_j)^2}
\end{equation}
where $(x_{0,i}, y_{0,i})$ are the fitted Gaussian centres of the emitters, and $(x'_j, y'_j)$ are the nearest points on the ideal grid.
Additionally, the radial discrepancy \((D_r)\) was calculated for each centre based on the offsets in \((x)\) and \((y)\), where

\begin{equation}
D_r = \min_j \sqrt{(x_0 - x'_j)^2 + (y_0 - y'_j)^2},
\end{equation}

This radial discrepancy ($D_r$) represents the distance between each observed centre and the ideal grid, providing a measure of how closely the observed centres align with the expected positions after accounting for the optimised offsets.

\begin{figure}[H]
\centering
\includegraphics[width=1\textwidth]{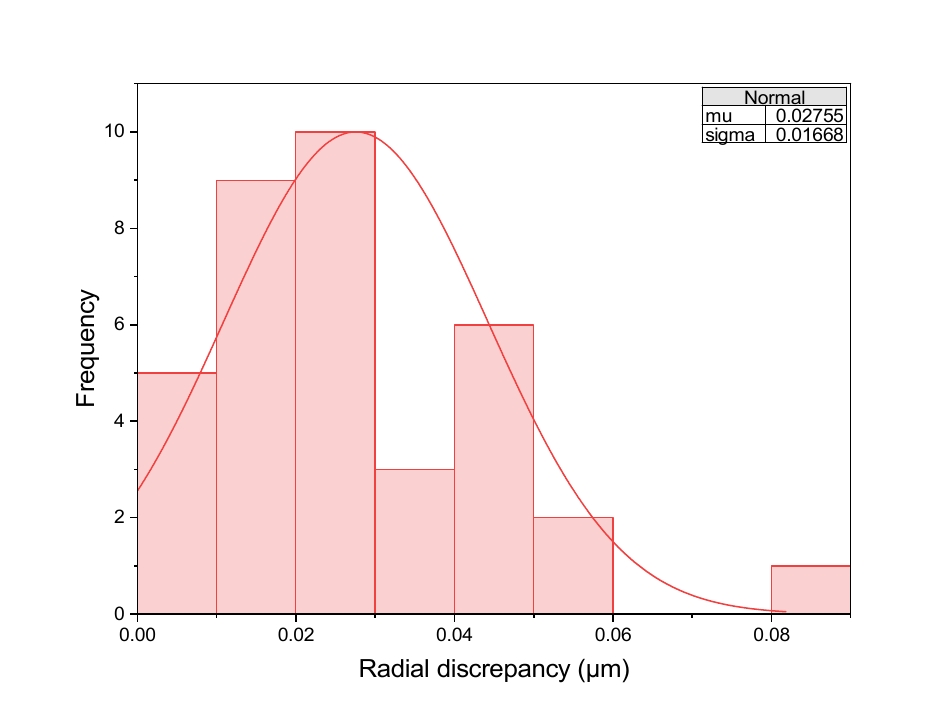}
\caption{\textbf{Histogram of radial discrepancies}. 
The histogram shows the distribution of radical discrepancies, with normal distribution fits overlaid. The fitted mean and standard deviation for the radial discrepancy are 27.6~nm and 16.9~nm, respectively.}
\label{Discrepency}
\end{figure}

Analysis of the optimised discrepancies revealed an average radial discrepancy of 27.6~nm. The sub-30 nm average discrepancy demonstrates the high precision of our combined implantation and activation process, closely aligning with the expected implantation positioning accuracy. Notably, the laser activation process does not alter the position of the implanted Sn atoms.

\subsection{Type II Sn}

\begin{figure}[H]
\centering
\includegraphics[width=1\textwidth]{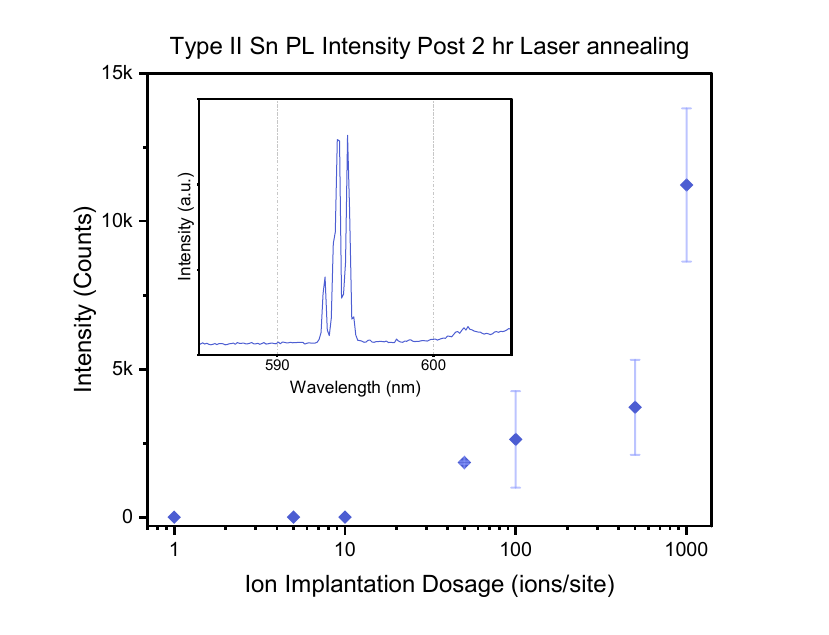}
\caption{\textbf{PL emission intensities of Type II Sn defects as a function of ion implantation dosage}.
PL emission intensities from post 2~h laser annealing Type II Sn were obtained by integrating the spectral window from 590 to 600~nm. A representative spectrum is shown in the inset.}\label{Type II Sn emission}
\end{figure}

Fig. \ref{Type II Sn emission} illustrates the relationship between implantation dosage and the PL intensities of Type II Sn defects, supplementing Fig. \ref{fig:1}. Higher     implantation dosages result in increased PL emission from Type II Sn, as more Sn ions are implanted into the diamond lattice.

To obtain a details of the multi-splitting of a Type II Sn ZPL, a high resolution spectrograph grating was employed. The emission is characterised by several sharp peaks within the wavelength range of approximately 593 nm to 596 nm, with the most prominent peak centred around 594.5 nm. The narrow linewidths of these peaks suggest well-defined electronic transitions. The line shape and intensities of each component peak within this emission remain stable, even with sub-second acquisition windows.

\begin{figure}[H]
\centering
\includegraphics[width=1\textwidth]{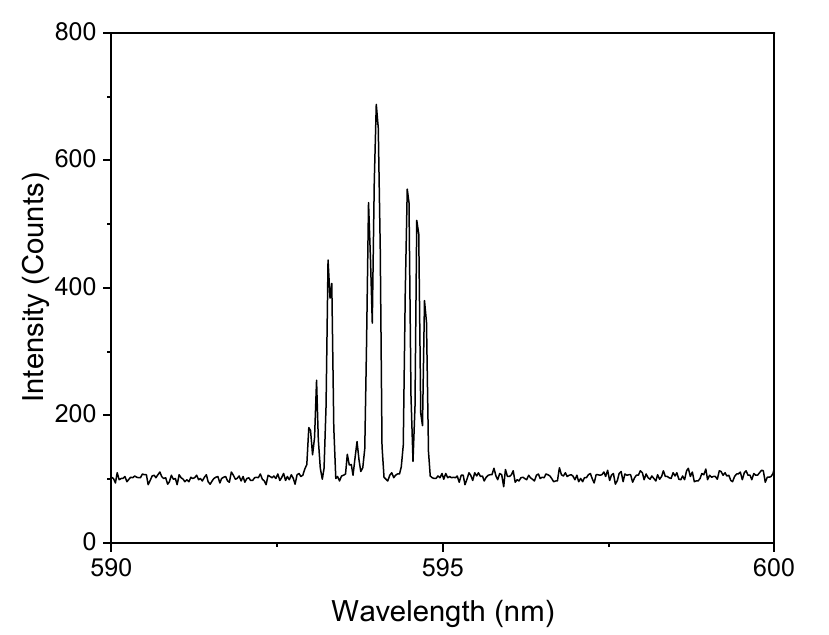}
\caption{\textbf{Spectral data of a typical Type II Sn defect, recorded using a high-resolution spectrograph grating}. 
}\label{1200blz Type II Sn}
\end{figure}

\begin{figure}[H]
\centering
\includegraphics[width=1\textwidth]{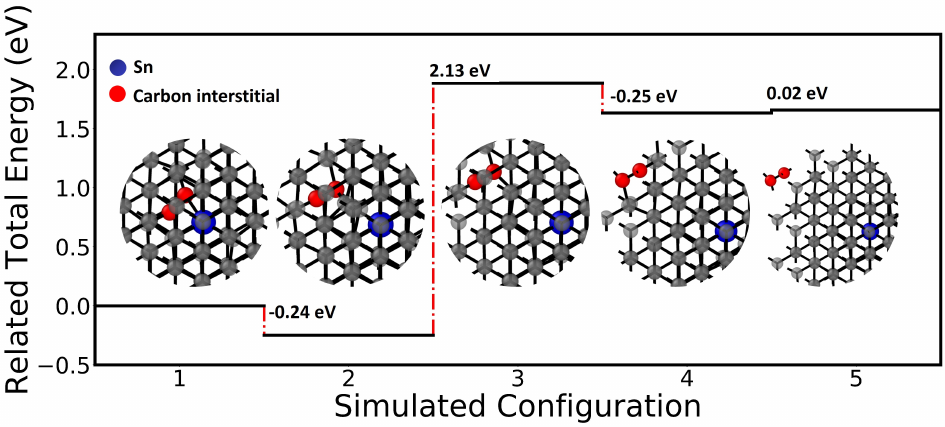}
\caption{\textbf{DFT simulation of Ci configurations relative to the implanted SnV defect in diamond.} The figure illustrates the total energy of the system for various simulated configurations of the Ci. In configuration 2, the Ci is in a stable state, occupying a second-nearest neighbour site to the SnV defect without recombining with the vacancy. The energy landscape shows an energy barrier that must be overcome to move the Ci away from the SnV complex, leading to the formation of a pure SnV$^-$ defect.}\label{DFT}
\end{figure}

Fig. \ref{DFT} presents the results of a DFT simulation, where Ci is oriented differently compared to the configuration discussed in the main text. The simulation explores various configurations of the Ci relative to the implanted SnV defect in the diamond lattice.

The DFT calculations reveal a stable configuration where the Ci resides at a second-nearest neighbour site to the SnV$^-$ defect, without recombining with the vacancy in the SnV$^-$ centre. This stable state is indicated by a local energy minimum in the second configuration, showing that the system can maintain this structure without spontaneous recombination. Energy barriers is presented for the Ci to move away from the SnV$^-$ centre, forming a pure SnV$^-$ emission.

\subsection{In-situ Monitoring}

\begin{figure}[H]
\centering
\includegraphics[width=1\textwidth]{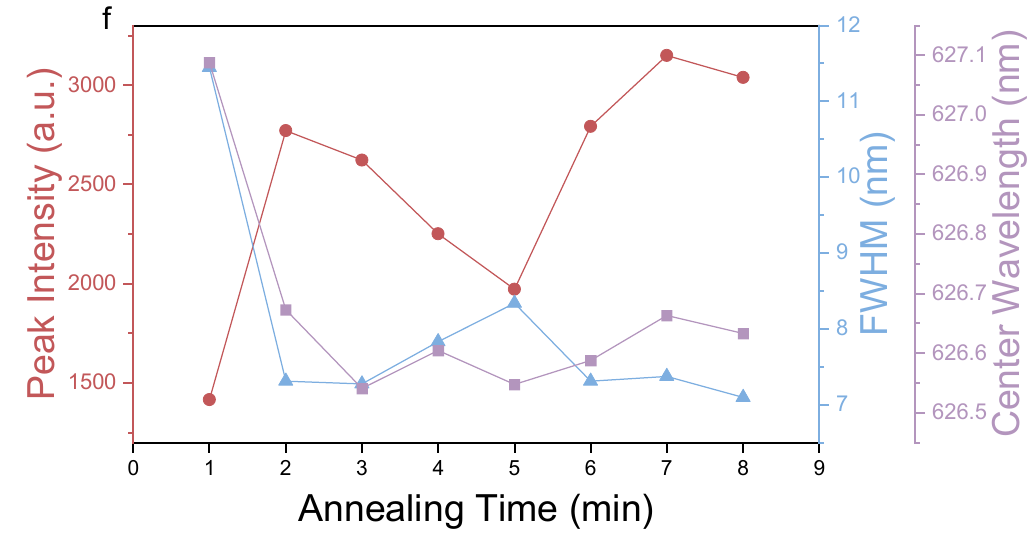}
\caption{\textbf{Spectrum tracking of the SnV$^-$ centre}. The ZPL of the SnV$^-$ centre, as shown in Fig.~\ref{fig:4}c of the main text, is monitored over time during laser annealing. A Lorentzian fit is applied to the ZPL, with the fitting parameters extracted and plotted as a function of annealing time.}
\label{peak change}
\end{figure}

As discussed in the main text, spectral shifts, structural changes, and intensity variations were observed during the laser annealing of the SnV$^-$ centre. Here, a Lorentzian function was fitted to the ZPL of the SnV$^-$ centre, and the fitting parameters were extracted. The peak intensities, determined by integrating the fitted Lorentzian peaks, vary with different annealing times, suggesting that the movement of Cis around the SnV$^-$ centre affects the degree of PL emission quenching. A correlation between peak intensity and full width at half maximum (FWHM) was observed: as the ZPL broadens, the SnV$^-$ emission intensity decreases, indicating that the presence of a Ci suppresses the SnV$^-$ emission. Notably, aside from the significant changes observed between 1 and 2 minutes—where both peak intensity and FWHM vary the most—the ZPL's central wavelength remains relatively stable until deactivation occurs.

\begin{figure}[H]
\centering
\includegraphics[width=1\textwidth]{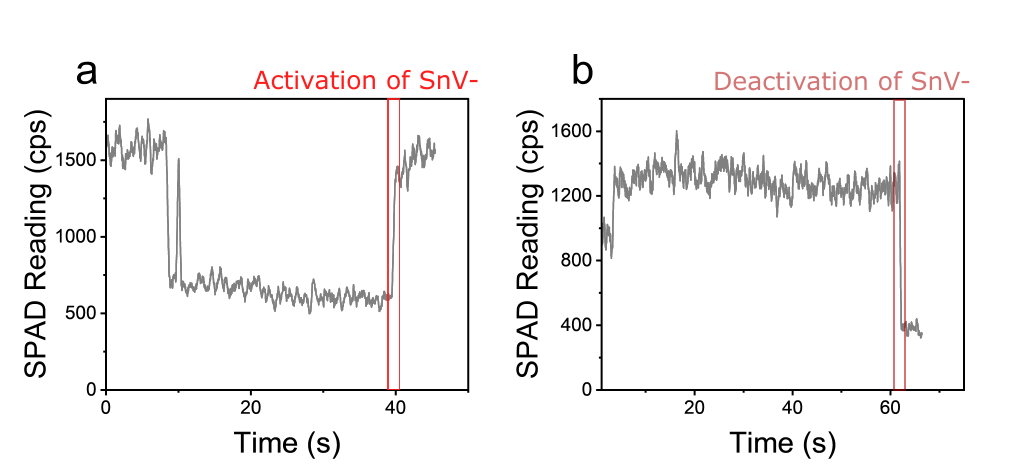}
\caption{\textbf{SPAD monitoring during laser annealing}. \textbf{a} shows a SPAD trace capturing the activation of a SnV$^-$ centre during laser annealing, while \textbf{b} presents the corresponding trace for the deactivation of a SnV$^-$ centre.}
\label{SPAD trace}
\end{figure}


In addition to spectral monitoring, in-situ tracking of the laser annealing process can be performed using fluorescence feedback loop via SPAD (Single-Photon Avalanche Diode) traces. SPAD readings with a binning time of 20 ms were recorded for emitters under fs-laser annealing, while simultaneously being excited by a 532~nm continuous-wave (CW) laser using a home-built confocal microscope. The PL emission signal from the emitter was collected within the $615-625$~nm range, where the SnV$^-$ emission is expected.
Fig.~\ref{SPAD trace}a shows the SPAD trace during the 4-5 minute interval in Fig.~\ref{fig:4} a of the main text. After a further 40 seconds of annealing from the 4-minute mark, a sharp increase in SPAD counts within the 615-625 nm emission window indicates the activation of a potential SnV$^-$ centre. The identity of the emitter was further confirmed to be SnV, as described in the main text. The initial decrease observed in the first 10 seconds of annealing, followed by a sharp spark, is not yet fully understood.

Fig.~\ref{SPAD trace}b displays the SPAD trace during the 3-4 minute interval in Fig.~\ref{fig:4} b of the main text. After an additional 60 seconds of annealing from the 3-minute mark, the SnV$^-$ emission deactivated, evidenced by a sharp drop in SPAD counts within the $615-625$~nm emission window.

\subsection{Laser activation of negatively charged silicon vacancy centres}

\begin{figure}[H]
\centering
\includegraphics[width=1\textwidth]{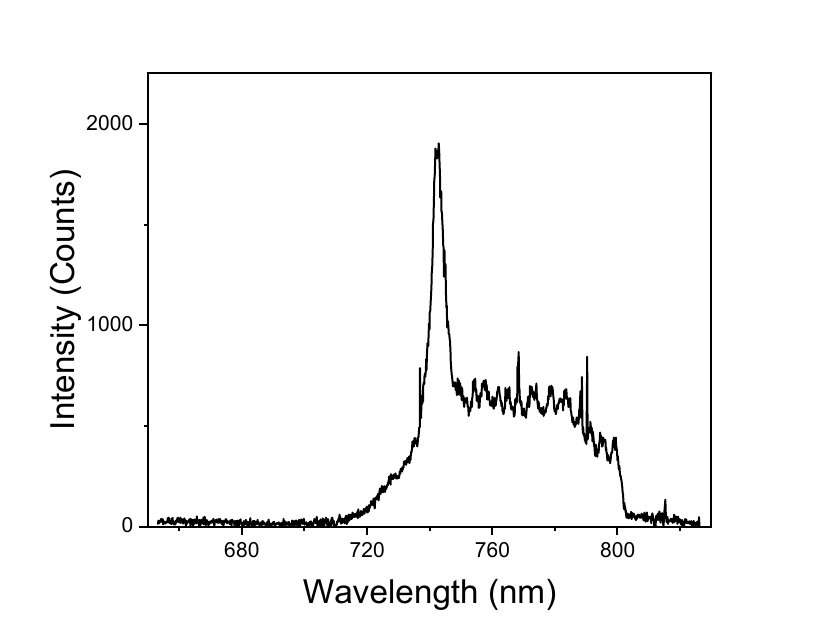}
\caption{\textbf{Emission spectrum of the SiV centre.} The spectrum shows a prominent ZPL at 737 nm, confirming the activation of the SiV centre. The oscillations observed in the peaks at higher wavelengths are due to etaloning effects.}
\label{SiVspec}
\end{figure}

This femtosecond laser activation technique can also be applied to other group-IV centres. In this study, we performed laser annealing on an ion-implanted array of 256 Si ions per site, following Poissonian statistics, with a separation of 10 \textmu m between sites. The sample was equipped with alignment marks in the x and y directions, corresponding to the implantation sites. This allowed precise laser focusing and site-by-site annealing with in-situ tracking via a fluorescence feedback loop.
The activation laser was turned off when a sharp increase in the SPAD counts was observed, indicating the activation of an emitter. To identify the type of emitters formed, corresponding spectra were recorded after laser activation. Fig.~\ref{SiVspec} shows a typical background-subtracted emission spectrum of an emitter from the activated array, with a sharp ZPL at 737~nm, characteristic of SiV$^-$\cite{Muller2014}, confirming the successful activation of SiV$^-$ centres using femtosecond laser pulses.

\begin{figure}[H]
\centering
\includegraphics[width=1\textwidth]{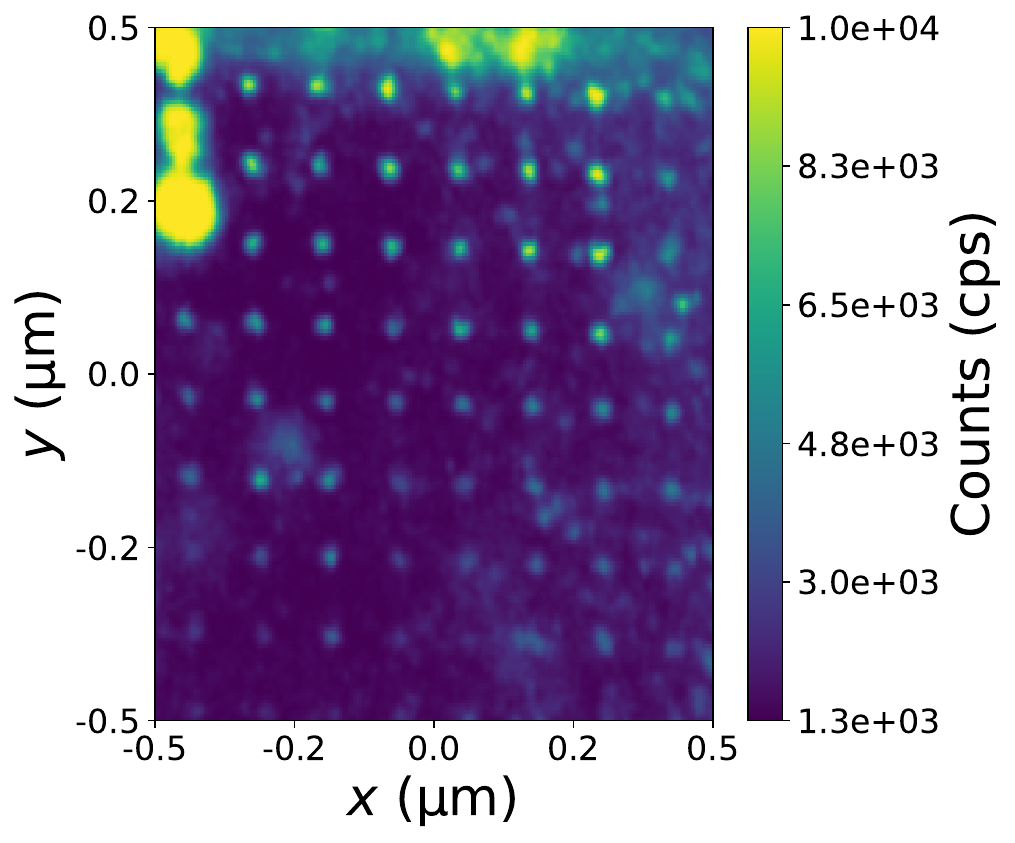}
\caption{\textbf{PL image of a post-laser activation Si-implanted array}.
2D PL maps of a ion implanted with Poissonian statistics 256 Si ions in each site, laser activation performed site by site with different annealing time and pulse energy.}\label{SiVPL}
\end{figure}

\bibliography{main}

\end{document}